\RequirePackage{fix-cm}
\documentclass[smallcondensed]{svjour3}     % onecolumn (ditto)
\smartqed  % flush right qed marks, e.g. at end of proof
\usepackage{graphicx}
\usepackage{amssymb}
\usepackage[mathscr]{eucal}
\usepackage{graphicx}

\usepackage{amsmath}
\usepackage{mathtools}

\usepackage[rightcaption]{sidecap}

\usepackage{graphicx} %package to manage images
\graphicspath{ {images/} }

\usepackage{mathptmx}      % use Times fonts if available on your TeX system
\usepackage{latexsym}
\usepackage{booktabs}
\usepackage{multirow}
\usepackage{tabularx}
\usepackage{caption}
\usepackage{latexsym}

 \usepackage{mathptmx}      % use Times fonts if available on your TeX system
\usepackage{latexsym}
\usepackage{booktabs}
\usepackage{multirow}
\usepackage{tabularx}
\usepackage{caption}
\begin{document}

\title{Periodic solution for strongly nonlinear 
oscillators by  He's new amplitude-frequency relationship
%	Application of He's  new amplitude-frequency relationship to strongly nonlinear 
%oscillators
%\thanks{Grants or other notes
%about the article that should go on the front page should be
%placed here. General acknowledgments should be placed at the end of the article.}
}
%\subtitle{Do you have a subtitle?\\ If so, write it here}

\titlerunning{Periodic solution for strongly nonlinear 
	oscillators}        % if too long for running head

\author{O. Gonz\'alez-Gaxiola     % \and
   %     G. Chac\'on-Acosta \and J. A. Santiago%etc.
}

%\authorrunning{Short form of author list} % if too long for running head

\institute{O. Gonz\'alez-Gaxiola \at
	Departamento de Matem\'aticas Aplicadas y Sistemas, Universidad Aut\'onoma Metropolitana-Cuajimalpa. Vasco de Quiroga 4871, Santa Fe, Cuajimalpa, 05300, Mexico D.F., Mexico\\
	%Tel.: +123-45-678910\\
	%Fax: +123-45-678910\\
	\email{ogonzalez@correo.cua.uam.mx}           %  \\
	%         \emph{Present address:} of F. Author  %  if needed
%	\and
%	G. Chac\'on-Acosta \at
%	Departamento de Matem\'aticas Aplicadas y Sistemas, Universidad Aut\'onoma Metropolitana-Cuajimalpa. Vasco de Quiroga 4871, Santa Fe, Cuajimalpa, 05300, Mexico D.F., Mexico
%	\and
%	J. A. Santiago \at
%	Departamento de Matem\'aticas Aplicadas y Sistemas, Universidad Aut\'onoma Metropolitana-Cuajimalpa. Vasco de Quiroga 4871, Santa Fe, Cuajimalpa, 05300, Mexico D.F., Mexico
}

\date{Received: date / Accepted: date}
% The correct dates will be entered by the editor

\maketitle

\begin{abstract}
This paper applies He's new amplitude-frequency relationship recently established by Ji-Huan He (Int J Appl Comput Math 3 1557-1560, 2017) to study periodic solutions
of strongly nonlinear systems with odd nonlinearities. Some examples are given to illustrate
the effectiveness, ease and convenience of the method. In general, the results are valid
for small as well as large oscillation amplitude. The method can
be easily extended to other nonlinear systems with odd nonlinearities and can therefore be found widely applicable
in engineering and other science. The method used in this paper can be applied directly to highly nonlinear problems without any discretization, linearization or additional requirements. 

% \PACS{PACS code1 \and PACS code2 \and more}
\end{abstract}

\section{Introduction}
\label{s:Intro}
\noindent Nonlinear vibration arises everywhere in science,  engineering and other disciplines, since most phenomena in our world today, are essentially nonlinear and are described by nonlinear equations. It is very important in applications to have a version of the frequency (or period) to have a better understanding of the phenomena modeled through differential equations that contain terms with high nonlinearities, and a simple mathematical method is very useful for practical applications.\\
\noindent Recently many analytical methods have appeared to obtain the approximate solutions of nonlinear systems, such as the parameter-expansion method \cite{Moh},  the harmonic balance method \cite{Tang,Nay-1,Mic,Bel1}, the energy balance method \cite{Yil,Khan-1}, the Hamiltonian approach \cite{Yil-1,He-x0},  the use of special functions \cite{Zu-1,Zu-2}, the max-min approach \cite{He-x,Ze},  the variational iteration method \cite{Gan,He-1,He-2,He-3,Waz-1} and homotopy perturbation \cite{Bel-1,Bel-2,Gan-1,Gan-2,He-4,He-5,He-6}, and others.  An excellent study, in which many of these techniques can be found in detail to solve nonlinear problems of oscillatory type can be seen in \cite{He-8}.\\
\noindent Recently, In \cite{He-y0} an analytical approximate technique for large and small amplitudes oscillations of a class of conservative single degree-of-freedom systems with odd
non-linearity is proposed. In this study, we have applied new method to find the approximate
solutions of nonlinear differential equation governing strongly nonlinear oscillators
and have made a comparison with the exact solution. The most interesting features of the used method are its
simplicity and its excellent accuracy of both period and corresponding
periodic solution for the entire range of oscillation
amplitude. Finally, four examples are presented to describe
the solution methodology and to illustrate the usefulness and
effectiveness of the proposed technique.
%The method is very easy and straightforward.

\section{He's new amplitude-frequency relationship }
\label{Rel}

\noindent Consider a one-dimensional, nonlinear oscillator governed
by

\begin{equation}\label{Eq-1}
u''+f(u)=0,
\end{equation}
with the initial conditions
\begin{equation}\label{Ic}
u(0)=A,\;\; u'(0)=0.
\end{equation}
where a prime denotes differentiation with respect to $t$ and the nonlinear function $f(u)$ is odd, i.e. $f (-u)=-f (u)$
and satisfies $f (u)/u>0$ for $u\in [-A,A]$, $u\neq0$. It is obvious
that $u = 0$ is the equilibrium position. The system oscillates
between the symmetric bounds $-A$ and $A$. The period and corresponding
periodic solution are dependent on the oscillation amplitude $A$.\\
\noindent According to He's new amplitude-frequency formulation, the approximate frequency  as a function of $A$ can be obtained as follows \cite{He-y0}:
\begin{equation}\label{Eq-2}
\omega^{2}(A)=\frac{\sum_{i=1}^{N}\omega_{i}^{2}(A)}{N}
\end{equation}
with each $\omega_{i}^{2}(A)$  defined by
\begin{equation}\label{Eq-3}
\omega_{i}^{2}(A)=f'(u_{i})
\end{equation}
where $u_i$ are location points, $0<u_{i}<A$. Explicitly, $u_{i}=iA/N$ for every $i=1,2,\ldots, N-1.$\\ The simplest way to calculate the frequency is given by
\begin{equation}\label{Eq-4}
\omega^{2}(A)=f'(u_{i}),
\end{equation}
for some $0<u_{i}<A$. The accuracy, however, depends greatly upon the location point.\\
In Table 1 we present the criteria suggested by Ji-huan He in \cite{He-y0} for choosing a suitable location point $u_i$.
\begin{table}[h!]
	\begin{center}
		\begin{tabular}{cc}
			%	\toprule
			\cmidrule(r){1-2}
			Conditions & Location point for Eq. (\ref{Eq-4}) \\
			\midrule
			$uf''(u)<0$&$0<u_{i}<A/2$\\
			$uf''(u)>0$&$A/2 \leq  u_{i}<A$\\
						\bottomrule
		\end{tabular}
	\end{center}
	\caption{Criterion for choosing a location point }
	\label{Tab-1}   
\end{table}

\noindent Therefore, the analytical approximate frequency $\omega$ as a function of $A$ is
\begin{equation}\label{Eq-5}
	\omega_{app}(A)=\sqrt{f'(u_{i})}.
\end{equation}
From Eq. (\ref{Eq-5}) we obtain the following approximate periodic solution to (\ref{Eq-1})
\begin{equation}\label{Eq-6}
	u_{app}(t)=A\cos \left(\sqrt{f'(u_{i})}\cdot t\right).
\end{equation}

%\vspace{0.1in}

\section{Numerical examples}
\label{sn:Exas}
In this section, we will give four examples to illustrate the use and the efectiveness of the present approach.\\

\vspace{0.1in}

\begin{figure}[h!]
	\begin{center}
		\includegraphics[width=80mm, height=55mm, scale=1.0]{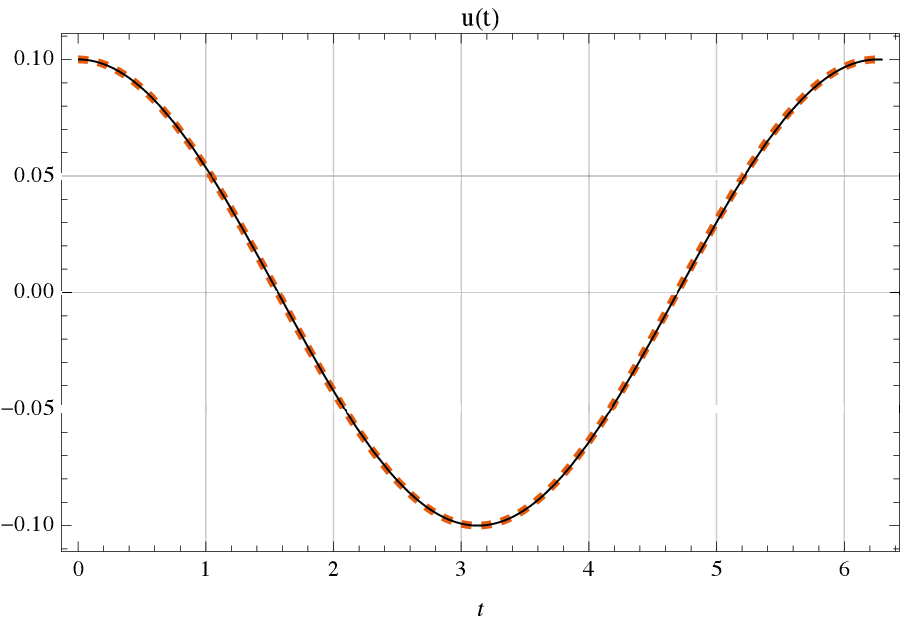}
	\end{center}
	\caption{Comparison of analytical approximation (dashed) and exact solution (black) for $A=1/10$ in example 1.\label{Gra-1}}
\end{figure}

\begin{figure}[h!]
	\begin{center}
		\includegraphics[width=80mm, height=55mm, scale=1.0]{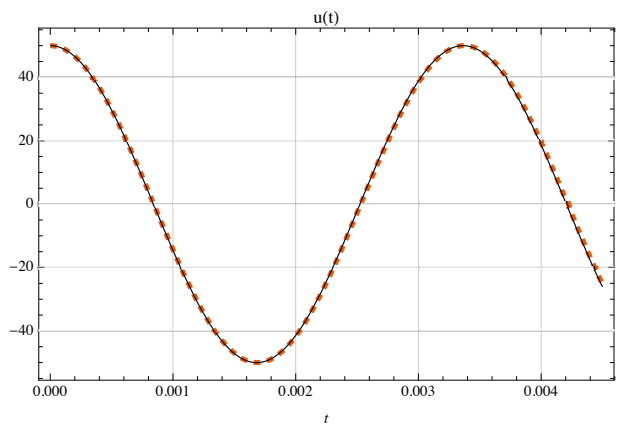}
	\end{center}
	\caption{Comparison of analytical approximation (dashed) and exact solution (black) for $A=50$ in example 1.\label{Gra-2}}
\end{figure}

\noindent {\bf Example 1}\\ Consider the cubic-quintic Duffing nonlinear oscillator, which
is modelled by the following second-order differential equation
\begin{equation}\label{Eq-7}
u''+u+u^{3}+u^{5}=0,
\end{equation}
with initial conditions
\begin{equation}\label{Eq-8}
u(0)=A,\;\; u'(0)=0.
\end{equation}
In the present example we have $f(u)=u+u^{3}+u^{5}$,  it is clear that $f$ is an odd function
and satisfies $f (u)/u>0$.\\
Calculating we have $f'(u)=1+3u^{2}+5u^{4}$ and $f''(u)=6u+20u^{3}$,  hence $uf''(u)>0$.
Now, considering the criterion given in Table \ref{Tab-1} we must take the location points $A/2\leq u_{i}<A$. If we take
$u_{i}=0.5772A$ and consider the proposed approach in Eq. (\ref{Eq-5}), one can assume for the frequency-amplitude formulation
\begin{equation}\label{Eq-9}
\omega_{app}(A)=\sqrt{1+3(0.5772)^{2}A^{2}+5(0.5772)^{4}A^{4}}.
\end{equation}
We, therefore, obtain the following periodic solution:
\begin{equation}\label{Eq-10}
	u_{app}(t)=A\cos \left(\sqrt{1+3(0.5772)^{2}A^{2}+5(0.5772)^{4}A^{4}}\cdot t\right)
\end{equation}
which has a high accuracy (see Figs. \ref{Gra-1}-\ref{Gra-2}).\\
The exact frequency for the present example is given by \cite{You}:
\begin{equation}\label{Eq-11}
\omega_{ex}(A) = \dfrac{2\pi}{\displaystyle \int_{0}^{\pi/2}\frac{4d\theta}{\sqrt{1+\frac{1}{2}(1+\sin^{2}\theta)A^{2}+\frac{1}{3}(1+\sin^{2}\theta+\sin^{4}\theta)A^{4}}}}.
\end{equation}
From Table \ref{Tab-2}, it can be observed that Eq. (\ref{Eq-9}) yield excellent analytical approximate periods for both small and large values of oscillation amplitude $A$.\\
\begin{table}[h!]
	\begin{center}
		\begin{tabular}{cccc}
			%	\toprule
			\cmidrule(r){1-4}
			$A$  & $\omega_{app}(A)$ Eq. (\ref{Eq-9}) &$\omega_{ex}(A)$ Eq. (\ref{Eq-11}) & Relative Error (\%)\\
			\midrule
			$1/1000$&$1.0000004997$&$1.0000003750$&$0.0000124\%$\\
			$1/100$&$1.0000499755$&$1.0000375023$&$0.0012401\%$\\
			$1/10$&$1.0050125835$&$1.0037729382$&$0.1234941\%$\\
			$10$&$75.171283755$&$75.177400632$&$0.0081365\%$\\
			$50$&$1863.0910920$&$1867.5739782$&$0.2400379\%$\\
			$100$&$7450.3513534$&$7468.8303066$&$0.2474142\%$\\
			$1000$&$744968.72043$&$746834.68847$&$0.2498502\%$\\
			\bottomrule
		\end{tabular}
	\end{center}
	\caption{Comparison between frequencies $\omega_{app}(A)$ and $\omega_{ex}(A)$ for different  values of $A$. }
	\label{Tab-2}   
\end{table}

\newpage

\noindent {\bf Example 2}\\ Consider the nonlinear oscillator
\begin{equation}\label{Eq-12}
u''+u+u^{5}=0,
\end{equation}
subject to the initial conditions
\begin{equation}\label{Eq-13}
u(0)=A,\;\; u'(0)=0.
\end{equation}
For this problem, 
$$f(u)=u+u^{5},$$
it is clear that $f$ is an odd function
and satisfies $f (u)/u>0$.\\
Derivating we have, $f'(u)=1+5u^{4}$ and $f''(u)=20u^{3}$,  hence $uf''(u)=20u^{4}>0$.
Therefore, considering the criterion given in Table \ref{Tab-1} we must take the location points $A/2\leq u_{i}<A$. If we take
$u_{i}=0.5779A$ and consider the proposed approach in Eq. (\ref{Eq-5}), one can assume for the frequency-amplitude formulation
\begin{equation}\label{Eq-14}
\omega_{app}(A)=\sqrt{1+5(0.5779)^{4}A^{4}}.
\end{equation}
The exact frequency for the present problem was established  in \cite{He-2} and is given by
\begin{equation}\label{Eq-15}
\omega_{ex}(A) = \frac{\pi\sqrt{A^{4}+3}}{2\sqrt{3}}\Bigg( \int_{0}^{\pi/2}\frac{1}{\sqrt{1+\Big(\frac{ A^{4}}{A^{4}+3}\Big)(\sin^{2}\theta+\sin^{4}\theta)}}\;d\theta\Bigg)^{-1}.
\end{equation}
To illustrate and verify accuracy of these approximate
analytical approach, a comparison of approximate  frequencies $\omega_{app}(A)$
for different values of amplitude $A$ and the exact frequencies $\omega_{ex}(A)$
is presented in Table \ref{Tab-3}. Note that the approximation is very accurate for small values and large values of $A$.
\begin{table}[h!]
	\begin{center}
		\begin{tabular}{cccc}
			%	\toprule
			\cmidrule(r){1-4}
			$A$  & $\omega_{app}(A)$ Eq. (\ref{Eq-14}) &$\omega_{ex}(A)$ Eq. (\ref{Eq-15}) & Relative Error (\%)\\
			\midrule
			$1/1000$&$1.0000000000$&$1.0000000000$&$0.0000000\%$\\
			$1/100$&$1.0000000028$&$1.0000000031$&$0.0000000\%$\\
			$1/10$&$1.0000278833$&$1.0000312493$&$0.0003365\%$\\
			$1$&$1.2480683052$&$1.2647077571$&$1.3156756\%$\\
			$10$&$74.684301857$&$74.690887847$&$0.0088176\%$\\
			$100$&$7467.7607379$&$7468.3420769$&$0.0077840\%$\\
			$500$&$186694.01678$&$186708.55006$&$0.0077839\%$\\
			$1000$&$746776.06710$&$746834.20022$&$0.0077839\%$\\
			$10000$&$7.467760*10^7$&$7.468342*10^7$&$0.0077839\%$\\
			\bottomrule
		\end{tabular}
	\end{center}
	\caption{Comparison between frequencies $\omega_{app}(A)$ and $\omega_{ex}(A)$ for different  values of $A$. }
	\label{Tab-3}   
\end{table}
From Table \ref{Tab-3} we can see that
\begin{equation}\label{Eq-16}
	\lim\limits_{A\to 0^{+}}\frac{\omega_{app}(A)}{\omega_{ex}(A)}=1 \quad \mbox{and}\quad 	\lim\limits_{A\to \infty}\frac{\omega_{app}(A)}{\omega_{ex}(A)}=0.999922.
\end{equation}
\noindent Considering the approximation for the frequency obtained in Eq. (\ref{Eq-14}) the approximate solution of Eq. (\ref{Eq-12})
becomes  
\begin{equation}\label{Eq-17}
u_{app}(t)=A\cos \left(\sqrt{1+5(0.5779)^{4}A^{4}}\cdot t\right).
\end{equation}
\noindent For this example we will not show graphs as we did in the previous example, because the high precision would not allow the distinction between them.

\vspace{0.1in}

%The obtained solution is of remarkable accuracy, as shown in Table 2.
\noindent {\bf Example 3}\\ Consider the cubic-quintic Duffing nonlinear oscillator, which
is modelled by the following second-order differential equation
\begin{equation}\label{Eq-18}
u''+\frac{1}{u}=0,
\end{equation}
with initial conditions
\begin{equation}\label{Eq-19}
u(0)=A,\;\; u'(0)=0.
\end{equation}
This is an important and interesting nonlinear differential equation  since it occurs in the modeling of certain phenomena in plasma physics \cite{La}.\\
The exact solution for Eq. (\ref{Eq-18}) as a function of $A$ was obtained in \cite{Mic-1} and this is
\begin{equation}\label{Eq-20}
\omega_{ex}(A) = 2\pi\Bigg[ 2\sqrt{2}A \int_{0}^{1}\frac{d s}{\sqrt{\ln(1/s)}}\Bigg]^{-1}.
\end{equation}
To use the method presented in the section \ref{Rel}, we will consider $f(u)=\frac{1}{u}$,  it is clear that $f$ is an odd function and satisfies $f (u)/u>0$.\\
Calculating, we get $f'(u)=-\frac{1}{u^2}$ and $f''(u)=\frac{2}{u^3}$,  hence $uf''(u)>0$.
Now, considering again the criterion given in Table \ref{Tab-1} we must take the location points $A/2\leq u_{i}<A$. If we take
$u_{i}=0.799A$ and consider the proposed approach in Eq. (\ref{Eq-5}), one can assume for the frequency-amplitude formulation
\begin{equation}\label{Eq-21}
\omega_{app}(A)=\sqrt{\frac{1}{(\frac{799}{1000})^{2}A^{2}}}=\frac{1000}{799A}.
\end{equation}

\begin{figure}[h!]
	\begin{center}
		\includegraphics[width=80mm, height=55mm, scale=1.0]{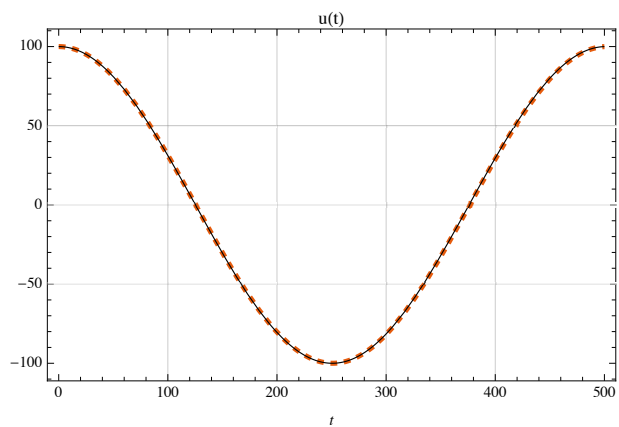}
	\end{center}
	\caption{Comparison of analytical approximation (dashed) and exact solution (black) for $A=100$ in example 3.\label{Gra-3}}
\end{figure}

\begin{table}[h!]
	\begin{center}
		\begin{tabular}{cccc}
			%	\toprule
			\cmidrule(r){1-4}
			$A$  & $\omega_{app}(A)$ Eq. (\ref{Eq-21}) &$\omega_{ex}(A)$ Eq. (\ref{Eq-20}) & Relative Error (\%)\\
			\midrule
			$1/1000$&$1251.5644556$&$1253.3141373$&$0.13960\%$\\
			$1/100$&$125.15644556$&$125.33141373$&$0.13960\%$\\
			$1/10$&$12.515644556$&$12.533141373$&$0.13960\%$\\
			$1$&$1.2515644556$&$1.2533141373$&$0.13960\%$\\
			$10$&$0.1251564456$&$0.1253314137$&$0.13960\%$\\
			$100$&$0.0125156446$&$0.0125331414$&$0.13960\%$\\
			$500$&$0.0025031289$&$0.0025066282$&$0.13960\%$\\
			$1000$&$0.0012515644$&$0.0012533141$&$0.13960\%$\\
			\bottomrule
		\end{tabular}
	\end{center}
	\caption{Comparison between frequencies $\omega_{app}(A)$ and $\omega_{ex}(A)$ for different  values of $A$. }
	\label{Tab-4}   
\end{table}
\begin{equation}\label{Eq-22}
\lim\limits_{A\to 0^{+}}\frac{\omega_{app}(A)}{\omega_{ex}(A)}=	\lim\limits_{A\to \infty}\frac{\omega_{app}(A)}{\omega_{ex}(A)}=0.9986.
\end{equation}
Finally, considering the approximation (\ref{Eq-21}), we have obtain the following periodic solution of the Eq. (\ref{Eq-18})
\begin{equation}\label{Eq-23}
u_{app}(t)=A\cos \left(\frac{1000}{799A}\cdot t\right).
\end{equation}
The obtained solution is of remarkable accuracy, as shown in Table \ref{Tab-4} and Fig. \ref{Gra-3}.\\

\vspace{0.1in}

\noindent {\bf Example 4}\\ As a last example, we consider the following nonlinear differential equation:
\begin{equation}\label{Eq-24}
u''+u+\frac{u}{\sqrt{1+u^2}}=0, \quad  u(0)=A,\;\; u'(0)=0.
\end{equation}
Which, $f(u)=u+\frac{u}{\sqrt{1+u^2}}$. Its derivatives are:
\begin{equation}\label{Eq-25}
f'(u)=1+\frac{1}{\sqrt{(1+u^2)^3}}, \quad f''(u)=-\frac{3u}{\sqrt{(1+u^2)^5}}.
\end{equation}
From Eq (\ref{Eq-25}) we have $uf''(u)<0$.
Considering the criterion given in Table \ref{Tab-1} we must take the location points $A<u_{i}<A/2$. If we take
$u_{i}=0.48A$ and consider the proposed approach in Eq. (\ref{Eq-5}), one can assume for the frequency-amplitude formulation
\begin{equation}\label{Eq-26}
\omega_{app}(A)=\sqrt{1+\frac{1}{(1+(\frac{48}{100})^2A^2)^{3/2}}}.
\end{equation}
The nonlinear oscillator described in Eq. (\ref{Eq-24}) is a conservative system. By integrating Eq. (\ref{Eq-24}) and using the initial conditions, we arrive at
\begin{equation}\label{Eq-27}
\omega_{ex}(A)=\frac{1}{2} \pi  \left(\int_0^{\frac{\pi }{2}} \frac{A\cos\theta}{\sqrt{A^{2}\cos^{2}\theta-2\big(\sqrt{1+A^2\sin^{2}\theta}-\sqrt{1+A^2}\big)}} d\theta\right)^{-1}
\end{equation}
By taking into account our approximation made through He's frequency-amplitude formulation Eq. (\ref{Eq-26})  and $\omega_{ex}(A)$ from Eq. (\ref{Eq-27})  we can calculate  the Table \ref{Tab-5} for small and large values of $A$.\\
\begin{table}[h!]
	\begin{center}
		\begin{tabular}{cccc}
			%	\toprule
			\cmidrule(r){1-4}
			$A$  & $\omega_{app}(A)$ Eq. (\ref{Eq-26}) &$\omega_{ex}(A)$ Eq. (\ref{Eq-27}) & Relative Error (\%)\\
			\midrule
			$1/1000$&$1.4142134402$&$1.4142134298$&$0.0000007\%$\\
			$1/100$&$1.4142013439$&$1.4142003049$&$0.0000734\%$\\
			$1/10$&$1.4129946662$&$1.4128952474$&$0.0070365\%$\\
			$1$&$1.3163234011$&$1.3273988465$&$0.8343720\%$\\
			$10$&$1.0042330178$&$1.0606052889$&$5.3151037\%$\\
			$100$&$1.0000045182$&$1.0063415277$&$0.6297076\%$\\
			$1000$&$1.0000000045$&$1.0006363862$&$0.0635976\%$\\
			$10000$&$1.0000000000$&$1.0000636597$&$0.0063655\%$\\
			\bottomrule
		\end{tabular}
	\end{center}
	\caption{Comparison between frequencies $\omega_{app}(A)$ and $\omega_{ex}(A)$ for different  values of $A$. }
	\label{Tab-5}   
\end{table}

\vspace{0.1in}

\begin{figure}[h!]
	\begin{center}
		\includegraphics[width=80mm, height=55mm, scale=1.0]{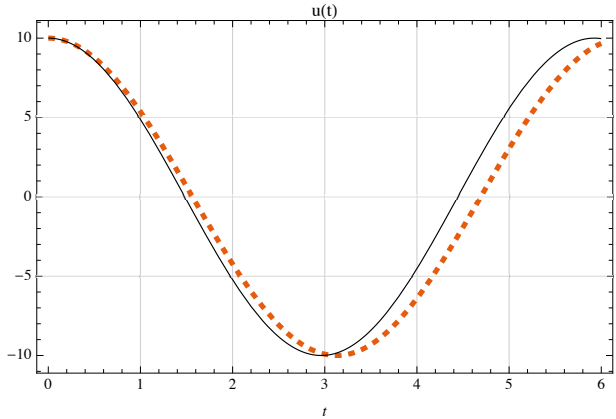}
	\end{center}
	\caption{Comparison of analytical approximation (dashed) and exact solution (black) for $A=10$ in example 4.\label{Gra-4}}
\end{figure}

\begin{figure}[h!]
	\begin{center}
		\includegraphics[width=80mm, height=55mm, scale=1.0]{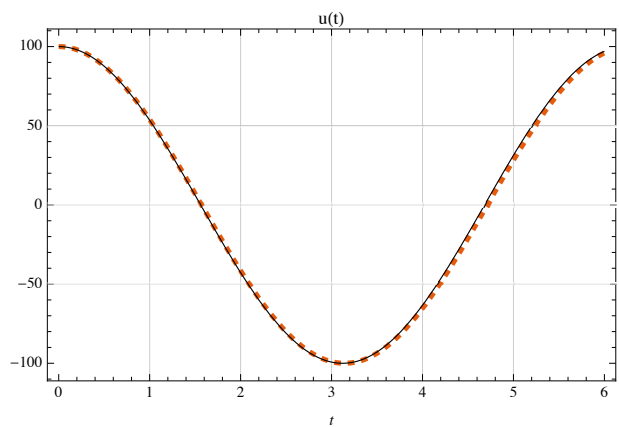}
	\end{center}
	\caption{Comparison of analytical approximation (dashed) and exact solution (black) for $A=100$ in example 4.\label{Gra-5}}
\end{figure}

\noindent Also, considering the approximation (\ref{Eq-26}), we have obtain the following periodic solution of the Eq. (\ref{Eq-24})
\begin{equation}\label{Eq-28}
u_{app}(t)=A\cos \left(\sqrt{1+\frac{1}{(1+(\frac{48}{100})^2A^2)^{3/2}}}\cdot t\right).
\end{equation}
The obtained solution is very acceptable accuracy, as shown in Fig. \ref{Gra-4} and Fig. \ref{Gra-5}. \\
We can conclude that formula (\ref{Eq-26}) is valid for the whole range
of values of amplitude of oscillation and its maximum
relative error is $5.3\%$ and this is obtained when
 $A = 10$. We can also see that, for very large or very small values of A, we have
\begin{equation}\label{Eq-29}
\lim\limits_{A\to 0^{+}}\frac{\omega_{app}(A)}{\omega_{ex}(A)}=	\lim\limits_{A\to \infty}\frac{\omega_{app}(A)}{\omega_{ex}(A)}=1.
\end{equation}


\vspace{0.5in}

\section{Conclusions}
\label{Con}
He's new amplitude-frequency relationship recently established by Ji-Huan He in \cite{He-y0}
 is proved to be a powerful mathematical tool for use in the search for periodic solutions of nonlinear oscillators. It is simple, straightforward and effective. Moreover the approximate analytical solutions are valid for small as well as large amplitudes of oscillation.\\
The new method applied in this paper is of potential and can be applied to other strongly nonlinear oscillators with more general restoring forces provided that they meet the requirements established in section \ref{Rel}.\\
Finally, four examples have been presented to illustrate
excellent accuracy of the analytical approximate periods and
the corresponding  periodic solutions. The technique is very
simple in principle, all numerical calculations have been made with the help of the software MATHEMATICA.

%\vspace{1.9in}

%\section*{Acknowledgments}

% Non-BibTeX users please use

\end{document}